\documentclass[11pt]{article}
\usepackage{amsmath}
\usepackage{amssymb}
\usepackage{amsthm}
\usepackage{xcolor}
\usepackage{caption}
\usepackage{calc}
\usepackage{hyperref}
\usepackage{graphicx}
\usepackage{wrapfig}
\usepackage{enumitem}
\usepackage{cite}
\usepackage{geometry}
\usepackage{algorithm}
\usepackage{dsfont}
\usepackage[noend]{algpseudocode}
\newcounter{algsubstate}
\renewcommand{\thealgsubstate}{\alph{algsubstate}}

\begin{document}

\title{Iterative execution of discrete and inverse discrete Fourier transforms with applications for signal denoising via sparsification}

 \author{H. Robert Frost$^{1}$}
\date{}
\maketitle
\begin{center}
\textit{
$^1$Department of Biomedical Data Science\\
Dartmouth College \\
Hanover, NH 03755, USA \\
rob.frost@dartmouth.edu
}
\end{center}

\begin{abstract}
We describe a family of iterative algorithms that involve the repeated execution of discrete and inverse discrete Fourier transforms. One interesting member of this family is motivated by the discrete Fourier transform uncertainty principle and involves the application of a sparsification operation to both the real domain and frequency domain data with convergence obtained when real domain sparsity hits a stable pattern. This sparsification variant has practical utility for signal denoising, in particular the recovery of a periodic spike signal in the presence of Gaussian noise. General convergence properties and denoising performance relative to existing methods are demonstrated using simulation studies. An R package implementing this technique and related resources can be found at \href{https://hrfrost.host.dartmouth.edu/IterativeFT}{https://hrfrost.host.dartmouth.edu/IterativeFT}.
\end{abstract}

\section{Problem statement}\label{sec:problem_statement}

We consider a class of iterative discrete Fourier transform \cite{Bracewell:2000ua} techniques described by Algorithm \eqref{alg:ft_it}. The structure of this algorithm is broadly motivated by iterative methods such as the alternating direction method of multipliers (ADMM) \cite{10.1561/2200000016} and expectation-maximization (EM) \cite{10.2307/2984875} and more specifically by iterative Fourier techniques such as the Gerchberg-Saxton (GS) algorithm \cite{Gerchberg72} and Fourier-based compressed sensing \cite{Stern:2015aa}. The family of algorithms we consider take a real-valued vector $\mathbf{x} \in \mathbb{R}^n$ as input and then repeatedly perform the following sequence of actions: 
\begin{itemize}[labelindent=5pt, topsep=4pt]\setlength{\itemsep}{2pt}
\item Execute a function $h(): \mathbb{R}^n \to \mathbb{R}^n$ on $\mathbf{x}$.
\item Perform a discrete Fourier transform, $dft()$, on the output of $h()$. The $j^{th}$ element of the complex-valued vector output by the discrete Fourier transform, $dft(\mathbf{x})[j]$, is defined as:
\begin{equation}\label{eqn:dft}
dft(\mathbf{x})[j] = \sum_{k=1}^{n} \mathbf{x}[k] e^{-i 2 \pi j k/n}
\end{equation}
\item Execute a function $g(): \mathbb{C}^n \to \mathbb{C}^n$  on the complex vector output by the discrete Fourier transform.
\item Transform the output of $g()$ back into the real domain via the inverse discrete Fourier transform, $dft^{-1}()$. The $j^{th}$ element of the real-valued vector output by the inverse discrete Fourier transform, $dft^{-1}(\mathbf{c})[j]$, is defined as:
\begin{equation}\label{eqn:inv_dft}
dft^{-1}(\mathbf{c})[j] = \frac{1}{n} \sum_{k=1}^{n} \mathbf{c}[k] e^{i 2 \pi j k/n}
\end{equation}
\end{itemize}
This iteration can expressed as:
\begin{equation}\label{eqn:iteration_rep}
\mathbf{x}_k = \begin{cases}
h(\mathbf{x}_k)\ & k = 0\\
h(dft^{-1}(g(dft(\mathbf{x}_{k-1})))) & k \geq 1\\
\end{cases}
\end{equation}
Convergence of the algorithm is determined by a function $c()$ that compares $\mathbf{x}_k$ to $\mathbf{x}_{k-1}$, i.e., the output of $h()$ on the current iteration to the version from the prior iteration. When convergence is obtained, the output from the last execution of $h()$ is returned. See Algorithm \eqref{alg:ft_it} for a detailed definition. For this general family of algorithms, a key question relates to what combinations of $h()$, $g()$, and $c()$ functions and constraints on $\mathbf{x}$ enable convergence. We are specifically interested in scenarios involving convergence to a value that is relevant for a specific data analysis application, e.g., a denoising or optimization problem.

\begin{align*}
h(dft^{-1}(g(dft(\mathbf{x}_{k-1})))) \\
\end{align*}


\begin{algorithm}
\caption{Iterative application of discrete Fourier and inverse Fourier transforms}
\label{alg:ft_it}
\hspace*{\algorithmicindent} \textbf{Inputs:}
\begin{itemize}
\setlength\itemsep{0em}
\item $\mathbf{x} \in \mathbb{R}^n$ \Comment{Input data}
\item $i_m$ \Comment{The maximum number of iterations}
\end{itemize}
\hspace*{\algorithmicindent} \textbf{Outputs:}
\begin{itemize}
\setlength\itemsep{0em}
\item $\mathbf{y} \in \mathbb{R}^n$ \Comment{Output data}
\item $i_c$ \Comment{Number of iterations completed}
\end{itemize}
\begin{algorithmic}[1]
\State $i = 1$ \Comment{Initialize iteration index}
\State $\mathbf{x}_0 = \mathbf{x}$ \Comment{Initialize $\mathbf{x}_i$}
\While {$i \leq i_m$}
\State $\mathbf{x}^*_i = h(\mathbf{x}_{i-1})$  \Comment{Apply function $h(): \mathbb{R}^n \to \mathbb{R}^n$ to $\mathbf{x}_{i-1}$}
\If{$c(\mathbf{x}^*_i, \mathbf{x}^*_{i-1})$}  \Comment{Check for convergence using indicator function $c()$ with domain $\{ \mathbb{R}^n, \mathbb{R}^n \}$}
  \State break \Comment{If convergence conditions met, stop the iteration}
\EndIf
\State $\mathbf{w}_i = \textit{dft}(\mathbf{x}^*_i)$ \Comment{Compute discrete Fourier transform of $\mathbf{x}^*_i$}
\State $\mathbf{w}^*_i = g(\mathbf{w}_i)$  \Comment{Apply function $g(): \mathbb{C}^n \to \mathbb{C}^n$ to complex vector $\mathbf{w}_i$}
\State $\mathbf{x}_{i} = \textit{dft}^{-1}(\mathbf{w}^*_i)$ \Comment{Compute inverse discrete Fourier transform of $\mathbf{w}^*_i$}
\State $i=i+1$ \Comment{Increment iteration index}
\EndWhile
\Return $(\mathbf{x}^*_i,i)$ \Comment{Return the output of the final execution of $h()$}
\end{algorithmic}
\end{algorithm}

\subsection{Trivial cases}\label{sec:trivial}

If both $h()$ and $g()$ are the identity function, then convergence occurs after a single iteration and the entire algorithm operates as the identity function, 
i.e., $h(dft^{-1}(g(dft(\mathbf{x}))) = dft^{-1}(dft(\mathbf{x})) = \mathbf{x}$. 
Similarly, if just one of $h()$ or $g()$ is the identity function, then the algorithm simplifies to the repeated execution of the non-identity function, 
e.g., if $g()$ is the identity function then $h(dft^{-1}(g(dft(\mathbf{x}))) = h(dft^{-1}(dft(\mathbf{x}))) = h(\mathbf{x})$. 
In general, we will assume that neither $h()$ nor $g()$ are the identity function and that the number of iterations until convergence, $i_c$, is a function of $\mathbf{x}$, i.e., if $\mathbf{x}$ is a random vector, then $i_c$ is a random variable. 

\subsection{Generalizations}\label{sec:generalizations}

A number of generalizations of Algorithm \eqref{alg:ft_it} are possible:
\begin{enumerate}[labelindent=5pt, topsep=4pt]\setlength{\itemsep}{2pt}
\item \textbf{Matrix-valued input}: Instead of accepting a vector $\mathbf{x} \in \mathbb{R}^{n}$ as input, the algorithm could accept a matrix $\mathbf{X} \in \mathbb{R}^{n \times m}$ (or higher-dimensional array) with $\textit{dft}()$ and $\textit{dft}^{-1}()$ replaced by two-dimensional counterparts.
\item \textbf{Complex-valued input}: Instead of just real values, elements of the input could be allowed to take complex or hypercomplex (e.g., quaternion or octonion) values with a correponding change to the complex, or hypercomplex, discrete Fourier transform.
\item \textbf{Alternative invertable discrete transform}: The discrete Fourier transform could be replaced by another invertable discrete transform, e.g., discrete wavelet transform \cite{10.5555/1525499}. More broadly, a similar approach could be used with any invertable discrete function.
\end{enumerate}
We explore the first generalization in this paper (see Section \ref{sec:matrix_denoising}) but restrict our interest to the real-valued discrete Fourier transform and leave the other generalizations to future work.

\section{Related techniques}\label{sec:related_techniques}

Algorithm \eqref{alg:ft_it} is related to both standard discrete Fourier analysis techniques, the iterative Gerchberg-Saxton (GS) algorithm \cite{Gerchberg72}, Fourier-based compressed sensing \cite{Stern:2015aa}, and iterative algorithms that alternate between dual problem representations. Although a detailed comparison is not possible without defining $h()$, $g()$, and $c()$, some general observations are possible.

\subsection{Relationship to standard discrete Fourier analysis}\label{sec:fourier_relationship}

The discrete Fourier transform is widely used in many areas of data analysis with common applications including signal processing \cite{Sundararajan:aa}, image analysis \cite{9243510}, and efficient evaluation of convolutions \cite{NEURIPS2020_2fd5d41e}. The standard structure for these applications is the transformation of real-valued data (e.g, time-valued signal or image) into the frequency domain via a discrete Fourier transform, computation on the frequency domain representation, and final transformation back into the time/location domain via an inverse discrete Fourier transform. For example, to preserve specific frequencies in a signal, the original data can be transformed into the frequency domain via a discrete Fourier transform, the coefficients for non-desired frequencies set to zero, and a filtered time domain signal generated via an inverse transformation. A similar structure is found in the application of other invertable discrete transforms. A key difference between this type of application and the approach discussed in this paper relates to number of executions of the forward and inverse transforms and the nature of the functions applied to the data in each domain, i.e., $h()$ and $g()$. Specifically, standard discrete Fourier analysis involves just the single application of forward and inverse transforms separated by execution of a single function. In contrast, the method described by Algorithm \eqref{alg:ft_it} involves the repeated execution of forward and inverse transforms interleaved by the execution of $h()$ and $g()$ functions until the desired convergence conditions are obtained.

\subsection{Relationship to Gerchberg-Saxton (GS) algorithm}\label{sec:GS_relationship}

The Gerchberg-Saxton (GS) algorithm \cite{Gerchberg72} is an interative Fourier-based technique that was developed to estimate phase information for electron microscopy data. In the case of electron microscopy data, only the amplitudes of the image and Fourier transform of the image (i.e., the diffraction plane) are measured and the phase information must be estimated. The GS method performs this estimation by initializing the phase information to random values and then iteratively applying forward and inverse discrete Fourier transforms on the image or diffraction plane amplitude data normalized using the most recent estimated phase information and then updating the phase estimates using the known amplitudes. The method terminates when the error between the reconstructed and measured amplitudes falls below a specific threshold. As detailed by Fienup \cite{Fienup:82}, the GS approach can be generalized beyond the phase estimation problem to a larger class of techniques, which Fienup calls error-reduction algorithms, that apply to problems where partial constraints are known in two domains coupled by forward and inverse transforms.

While the GS method iteratively applies forward and inverse discrete Fourier transforms, it is functionally distinct from the class of methods considered in this paper. Specifically, we are considering methods that only accept just a single input (e.g., real domain data) and iteratively apply specified functions to the data and Fourier transformation of the data. In contrast, the GS method, and more general error-reduction techniques, accept two inputs that represent partial constraints in both domains (e.g., the phaseless amplitudes of a dataset and the amplitudes of the Fourier transform of that data) and then use the iteration to estimate the missing information (e.g., the phase).

\subsection{Relationship to Fourier-based compressed sensing}\label{sec:CS_relationship}

Another related set of algorithms are Fourier-based compressed sensing (CS) methods. An example of this type of CS methods for NMR data is detailed by Stern et al. \cite{Stern:2015aa} and applies the following steps iteratively to measured real data: 1) apply discrete Fourier transform, 2) apply hard thresholding on the frequency domain representation, 3) convert back to the real domain via an inverse discrete Fourier transform, and 4) replace all observed entries with the measured data. While Fourier-based CS techniques do involve repeated execution of forward and inverse descrete Fourier transforms they only apply hard thresholding in the frequency domain so will not converge to a sparse real domain solution.

\subsection{Relationship to other iterative algorithms}\label{sec:iterative_relationship}

Algorithm \eqref{alg:ft_it} shares features with a range of methods (e.g., ADMM \cite{10.1561/2200000016}, Dykstra's algorithm \cite{02cb055f-7013-3403-b561-c462e7d6404a}, and EM \cite{10.2307/2984875}) that alternate between coupled representations of a problem on each iteration. For ADMM, an optimization problem is solved by alternating between 1) estimating the value of the primal variable that minimizes the Lagrangian with the Lagrangian variables fixed and 2) updating the Lagrangian or dual variables using the most recent primal variable value. Dykstra's algorithm is also used to solve optimization problems and, for many scenarios, is exactly equivalent to ADMM \cite{10.5555/3294771.3294821}. For EM, a likelihood maximization problem is solved by alternatively 1) finding the probability distribution of latent variables that maximizes the expected likelihood using fixed parameter values and 2) finding parameter values that maximize the expected likihood given the most recent latent variable distribution.  ADMM, Dykstra's algorithm and EM can all therefore be viewed as techniques that alternatively update different parameter subsets of a common function. In contrast, Algorithm \eqref{alg:ft_it} alternates between two different functions, $h()$ and $g()$, that are applied to the same data before and after an invertable discrete transform. Thus, while Algorithm \eqref{alg:ft_it} is broadly similar to these techniques, it cannot be directly mapped to these methods. We are not aware of existing iterative algorithms that are equivalent to Algorithm \eqref{alg:ft_it}. 

\section{Iterative convergence under sparsification}\label{sec:convergence}

An interesting subclass of the general iterative method detailed in Algorithm \ref{alg:ft_it} involves the use of sparsification functions for both $h()$ and $g()$ with $c()$ identifying convergence when a stable sparsity pattern is achieved in the output of $h()$. Iterating between real domain and frequency domain sparsification is motivated by the discrete Fourier transform uncertainty principal \cite{doi:10.1137/0149053}, which constrains the total number of zero values in the real domain data and frequency domain representation generated by a discrete Fourier transform. Attempts to induce sparsity in one domain will reduce sparsity in the other domain with the implication that setting both $h()$ and $g()$ to sparsification functions will not simply result in the generation of a vector of all 0 elements. Instead, for scenarios where the iterative algorithm converages, the solution will represent a stable compromise between real and frequency domain sparsity.

We can explore the general convergence properties of Algorithm \ref{alg:ft_it} using sparsification functions for $h()$ and $g()$ via the following simulation design:

\begin{itemize}[labelindent=5pt, topsep=4pt]\setlength{\itemsep}{2pt}
\item Set $\mathbf{x}$ to a length $n$ vector of $\mathcal{N}(0,1)$ random variables. 
\item Define $h()$ to generate a sparse version of $\mathbf{x}$ where the proportion $p$ of elements with the smallest absolute values are set to 0.
\item Define $g()$ to generate a sparse version of $\mathbf{w}$ where the proportion $p$ of complex coefficients with the smallest magnitudes are set to $0 + 0i$.
\item Define $c()$ to identify convergence when the indices of 0 values in the output of $h()$ are identical on two sequential iterations.
\item The discrete and inverse discrete Fourier transforms are realized using the Fast Fourier Transform.
\end{itemize}

\begin{figure}[t]
\begin{center}
\includegraphics[width=0.65\textwidth]{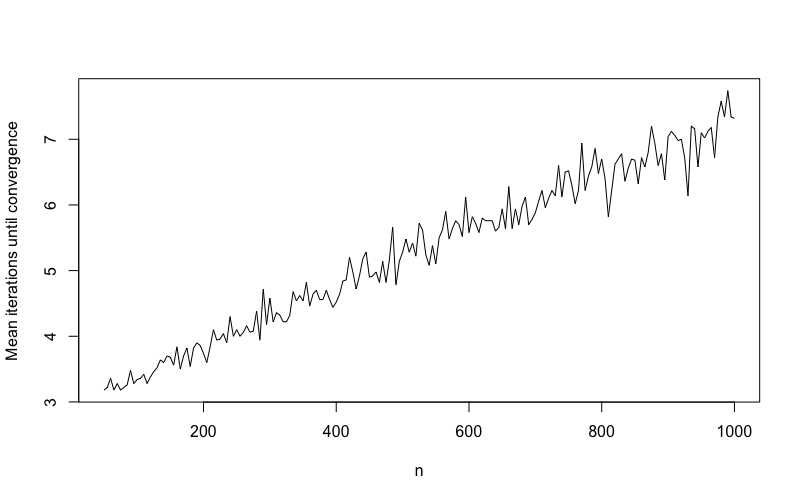}
\end{center}
\vspace{-1cm}
\caption{Mean iterations until convergence for random length $n$ vectors of $\mathcal{N}(0,1)$ random variables and $h()$ and $g()$ functions that rank order the elements according to absolute value or magnitude and set the bottom 50\% to 0. Convergence is based on repeating the same pattern of sparsity after execution of $h()$ on two sequential iterations.}
\label{fig:convergence_vs_n}
\end{figure}

\begin{figure}[h]
\begin{center}
\includegraphics[width=0.75\textwidth]{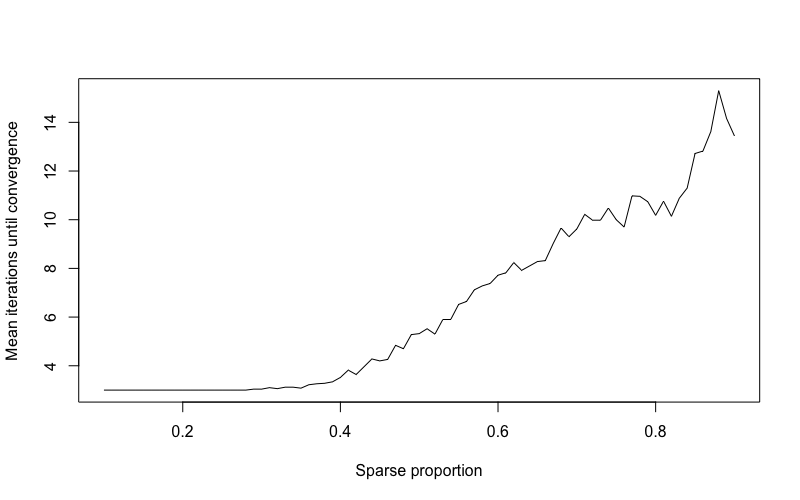}
\end{center}
\caption{Mean iterations until convergence for random length 500 vectors of $\mathcal{N}(0,1)$ random variables. For these results $h()$, $g()$, and $c()$ have similar definitions as detailed for Figure \ref{fig:convergence_vs_n}. In this case, $n$ was fixed at 500 and the sparsity proportion varied between 0.1 and 0.9.}
\label{fig:convergence_vs_p}
\end{figure}

\noindent Both general and sparsification versions of Algorithm \ref{alg:ft_it} are supported by the IterativeFT R package available at
\href{https://hrfrost.host.dartmouth.edu/IterativeFT}{https://hrfrost.host.dartmouth.edu/IterativeFT}.
Following this design, we applied the algorithm to 50 simulated $\mathbf{x}$ vectors for each distinct $n$ value in the range from 50 to 1,000 using the sparse proportion of $p=0.5$ and the maximum number of iterations $i_m=50$. Figure \ref{fig:convergence_vs_n} below displays the mean number of iterations until convergence as a function of $n$. Figure \ref{fig:convergence_vs_p} illustrates the results from a similar simulation that used a fixed $n$ of 500 and sparse proportion value ranging from $0.1$ to $0.9$.
For all of the tests visualized in Figures \ref{fig:convergence_vs_n} and \ref{fig:convergence_vs_p}, the algorithm converged to a stable pattern of sparsity in $\mathbf{x}$. Not surprisingly, the number of iterations required to achieve a stable sparsity pattern increased with the growth in either $n$ or $p$. 
If $h()$ and $g()$ are changed to set all elements with absolute value or magnitude below the mean to 0 (see simulation design in Section \ref{sec:denoising}), the relationship between mean iterations until convergence and $n$ is similar to that shown in Figure \ref{fig:convergence_vs_n}. Changing the generative model for $\mathbf{x}$ to include a non-random periodic signal (e.g., sinusodial signal or spike signal) also generates similar convergence results.

\clearpage

\section{Theoretical convergence}\label{sec:convergence_proof}


The sparsification version of Algorithm \ref{alg:ft_it} seeks to minimize the following function of the output of $h()$ on sequential iterations ($\mathbf{x}^*_i$ and $\mathbf{x}^*_{i+1}$ using the notation in Algorithm \ref{alg:ft_it}):
\begin{equation}\label{eqn:objective_function}
F(\mathbf{x}^*_{i}, \mathbf{x}^*_{i+1}) = \| \mathds{1}(\mathbf{x}^*_{i}) - \mathds{1}(\mathbf{x}^*_{i+1}) \|_0
\end{equation}
\noindent where $\mathds{1}()$ is a support indicator function (i.e., $\mathds{1}(a) = 1$ if $a \neq 0$ else $\mathds{1}(a) = 0$) with array-valued input and output and $||x||_0$ is the $\ell^0$ norm (i.e., number of non-zero elements). The function $F()$ thus outputs the number of differences between the supports of $\mathbf{x}^*_i$ and $\mathbf{x}^*_{i+1}$ and Algorithm \ref{alg:ft_it} will converge when $F(\mathbf{x}^*_{i}, \mathbf{x}^*_{i+1}) = 0$, i.e., the support is stable across iterations. This equation can be rewritten in terms of just $\mathbf{x}^*_i$ with $h()$ and $g()$ replaced by the hard thresholding functions $H_{\lambda_r}()$ and $H_{\lambda_f}()$ with $\lambda_r$ and $\lambda_f$ the effective thresholding levels in the real and frequency domains:
\begin{equation}\label{eqn:objective_function_2}
F(\mathbf{x}^*_{i}) = \| \mathds{1}(\mathbf{x}^*_{i}) - \mathds{1}(H_{\lambda_r}(dft^{-1}(H_{\lambda_f}(dft(\mathbf{x}^*_{i}))) \|_0
\end{equation}

\noindent To prove covergence of the sparsification version of Algorithm \ref{alg:ft_it} it is therefore sufficient to show that the support of $\mathbf{x}^*_{k}$ becomes frozen for some finite $k > K$. This can be demonstrated using a three step procedure:

\begin{enumerate}
\item Specify the minimum $\ell_2$ change in $\mathbf{x}^*_i$ between iterations.
\item Define a continuous Lyapunov surrogate function $L()$ and show that it satisfies a sufficient decrease condition.
\item Specify the maximum number of support changes by iteration $k$ and show that there exists a $k=K$ beyond which no further support changes occur.
\end{enumerate}

\subsection{Minimum $\ell_2$ change in $\mathbf{x}^*_i$}

The support of $\mathbf{x}^*$ can only change between iteration $i$ and iteration $i+1$ if at least one element moves from being above the threshold $\lambda_r$ to below or vice versa. For an element $j$, this implies that either $\mathbf{x}_{i}[j] \geq \lambda_r$ and $\mathbf{x}_{i+1}[j] < \lambda_r$ or $\mathbf{x}_{i}[j] < \lambda_r$ and $\mathbf{x}_{i+1}[j] \geq \lambda_r$. For both of these cases, the thresholded element changes by at least $\lambda_r$ between iterations, i.e., goes from 0 to $\geq \lambda_r$ or from $\geq \lambda_r$ to 0. Specifically, the squared $\ell_2$ distance in that element after thresholding is bounded below by the square of the threshold: $\| \mathbf{x}^*_{i+1}[j] - \mathbf{x}^*_{i}[j] \|_2^2 \geq \lambda_r^2$.
This implies that if the objective function is greater than 0, i.e., support is not frozen, the total squared $\ell_2$ distance between the $\mathbf{x}^*$ values must be greater than $\lambda_r^2$:
\begin{equation} \label{eqn:min_change}
\|\mathbf{x}^*_{i+1} - \mathbf{x}^*_{i} \|_2^2 \geq \lambda_r^2
\end{equation}

\subsection{Sufficient decrease of Lyapunov function $L()$}

Equation \ref{eqn:min_change} can be leveraged to define a continuous Lyapunov function $L()$ that can be shown to decrease every iteration. 
%
Specifically, $L()$ will be defined by the squared $\ell_2$ norm of the difference between $\mathbf{x}^*_{i}$ and $\mathbf{x}_{i}$ (the value before thresholding):
\begin{equation}\label{eqn:lyapunov}
L(\mathbf{x}^*_{i}) = \| \mathbf{x}^*_{i} - \mathbf{x}_{i} \|_2^2 = \| \mathbf{x}^*_{i} - dft^{-1}(H_{\lambda_f}(dft(\mathbf{x}^*_{i}))) \|_2^2
\end{equation}

\noindent $L()$ can be shown to satisfy a sufficient descent condition (i.e., $L()$ decreases by at least an amount proportional to the squared $\ell_2$ difference in the $\mathbf{x}^*$ values each iteration):
\begin{equation}\label{eqn:sufficient_descent}
L(\mathbf{x}^*_{i+1}) + \alpha \| \mathbf{x}^*_{i+1}-\mathbf{x}^*_{i} \|_2^2 \leq L(\mathbf{x}^*_{i})
\end{equation}

\noindent Satisfaction of this condition can be shown as follows:

\begin{itemize}
\item Hard thresholding solves the following optimization problem: $\text{arg min}_{x} \left\{ \frac{1}{2} \|x - y\|_2^2 + \frac{\lambda^2}{2} \|x\|_0 \right\}$
\item Hard thresholding in the frequency domain, $\mathbf{w}^*_{i} = H_{\lambda_f}(dft(\mathbf{x}^*_{i}))$, therefore minimizes $\frac{1}{2} \| \mathbf{w}^*_{i} - dft(\mathbf{x}^*_{i})\|_2^2 + \frac{\lambda^2}{2} \| \mathbf{w}^*_{i} \|_0$ and thus provides a descent step $\Delta F_{freq} \geq a \| \mathbf{w}^*_{i} - dft(\mathbf{x}^*_{i}) \|_2^2$.
\item Hard thresholding in the real domain, $\mathbf{x}^*_{i+1} = H_{\lambda_r}(dft^{-1}(\mathbf{w}^*_{i}))$, therefore minimizes $\frac{1}{2} \| \mathbf{x}^*_{i+1} - dft^{-1}(\mathbf{w}^*_{i})\|_2^2 + \frac{\lambda^2}{2} \| \mathbf{x}^*_{i+1} \|_0$ and thus provides a descent step $\Delta F_{real} \geq b \| \mathbf{x}^*_{i+1} - dft^{-1}(\mathbf{w}^*_{i}) \|_2^2$.
\item The total descent in $L()$ one iteration (i.e., $L(\mathbf{x}^*_{i}) - L(\mathbf{x}^*_{i+1})$) is the sum of the frequency and real domain descents:
$ L(\mathbf{x}^*_{i}) - L(\mathbf{x}^*_{i+1}) = \Delta F_{real} + \Delta F_{freq} \geq a \| \mathbf{w}^*_{i} - dft(\mathbf{x}^*_{i}) \|_2^2 + b \| \mathbf{x}^*_{i+1} - dft^{-1}(\mathbf{w}^*_{i}) \|_2^2$.
\item By Parseval's Theorem \cite{Parseval1806}, i.e., the unitary property of the DFT, we know that $ \| \mathbf{w}^*_{i} - dft(\mathbf{x}^*_{i}) \|_2^2 = \| dft^{-1}(\mathbf{w}^*_{i}) - \mathbf{x}^*_{i} \|_2^2$. This can be used to rewrite the overall descent step inequality as $ L(\mathbf{x}^*_{i}) - L(\mathbf{x}^*_{i+1}) = \Delta F_{real} + \Delta F_{freq} \geq a \| dft^{-1}(\mathbf{w}^*_{i}) - \mathbf{x}^*_{i} \|_2^2 + b \| \mathbf{x}^*_{i+1} - dft^{-1}(\mathbf{w}^*_{i}) \|_2^2$

\item By the triangle inequality ($\|a\|_2^2 + \|b\|_2^2 \geq \frac{1}{2}\|a+b\|_2^2$) and simple algebra, this can be rewritten as 
$L(\mathbf{x}^*_{i+1}) + \alpha \| \mathbf{x}^*_{i+1} - \mathbf{x}^*_{i} \|_2^2 \leq L(\mathbf{x}^*_{i})$, which shows that $L()$ satisfies the sufficient descent condition.

\end{itemize}

\subsection{Maximum number of support changes by iteration $k$}

Equations \eqref{eqn:min_change} and \eqref{eqn:sufficient_descent} can be combined to generate:
\begin{equation}\label{eqn:sufficient_descent_lambda}
L(\mathbf{x}^*_{i+1}) + \alpha \lambda_r^2 \leq L(\mathbf{x}^*_{i})
\end{equation}

\noindent which implies that any change in the support decreases the value of $L()$ by at least $\alpha \lambda_r^2$. Expanding \eqref{eqn:sufficient_descent_lambda} to cover the range from the first iteration to iteration $k$ yields:
\begin{equation}\label{eqn:sufficient_descent_full}
L(\mathbf{x}^*_{k}) + N_s \alpha \lambda_r^2 \leq L(\mathbf{x}^*_{1})
\end{equation}

\noindent where $N_s$ is the total number of support changes up to iteration $k$. Because $L()$ is defined as a squared norm the value must be $\geq 0$. This allows \eqref{eqn:sufficient_descent_full} to be rewritten with $L(\mathbf{x}^*_{K})$ replaced by 0 to provide an upper bound on $N_s$:
\begin{equation}\label{eqn:N_upper_bound}
N_s \leq \frac{L(\mathbf{x}^*_{1})}{\alpha \lambda_r^2}
\end{equation}

\noindent Since $L(\mathbf{x}^*_{1})$ is finite and $\alpha \lambda_r^2 > 0$, $N_s$ must be finite, which implies that there is a finite iteration $k=K$ beyond which no further support changes are possible, satisfying the overall covergence condition of $\| \mathds{1}(\mathbf{x}^*_{i}) - \mathds{1}(\mathbf{x}^*_{i+1}) \|_0 = 0$.

\section{Detection of spike signals using iterative convergence}\label{sec:denoising}

To assess the pratical utility of a sparsification version of Algorithm \ref{alg:ft_it} for signal denoising, we applied the technique to data generated as either a one-dimensional vector $\mathbf{x}$ or two-dimensional matrix $\mathbf{X}$ containing a periodic spike signal with Guassian noise. For the one-dimensional cases, $h()$, $g()$, and $c()$ were specified as follows:

\begin{itemize}[labelindent=5pt, topsep=4pt]\setlength{\itemsep}{2pt}
\item Define $h()$ to generate a sparse $\mathbf{x}$ where all elements $|x_i| \leq 1/n\sum_{j=1}^n |x_j|$ are set to 0. 
\item Define $g()$ to generate a sparse $\mathbf{w}$ where all elements $|w_i| \leq 1/n\sum_{j=1}^n |w_j|$ are set to $0 + 0i$ (here the $||$ operation represents the magnitude of the complex number $w_j$). 
\item Define $c()$ to identify convergence when the indices of 0 values in the output of $h()$ are identical on two sequential iterations.
\end{itemize}

\noindent For the matrix case, $h()$, $g()$, and $c()$ can be executed on a vectorized version of $\mathbf{X}$. In the remainder of this paper, we will refer to the version of Algorithm \ref{alg:ft_it} that uses these $h()$, $g()$, and $c()$ functions as the IterativeFT method. 

\subsection{Detection of spike signal in vector-valued input}\label{sec:vector_denoising}

For the one-dimensional case, the input vector $\mathbf{x}$ was generated as the combination of a periodic spike signal $\mathbf{s}$ and Gaussian noise $\boldsymbol{\varepsilon}$, $\mathbf{x} = \mathbf{s} + \boldsymbol{\varepsilon}$, with:

\begin{itemize} [labelindent=5pt, topsep=4pt]\setlength{\itemsep}{2pt}
\item Elements $s_i, i \in (1,...,n)$ of $\mathbf{s}$ set to 0 for all $i \neq a\lambda$ and generated as $U(\alpha_{min},\alpha_{max})$ random variables for $i = a\lambda, a \in (1,...,b)$ where $b$ is the total number of cycles and $\lambda \geq 1$ is the period. 
\item Elements $\varepsilon_i$ of $\boldsymbol{\varepsilon}$ are generated as independent random variables with distribution $\mathcal{N}(0,\sigma^2)$.
\end{itemize}

\begin{figure}[h]
\begin{center}
\includegraphics[width=0.79\textwidth]{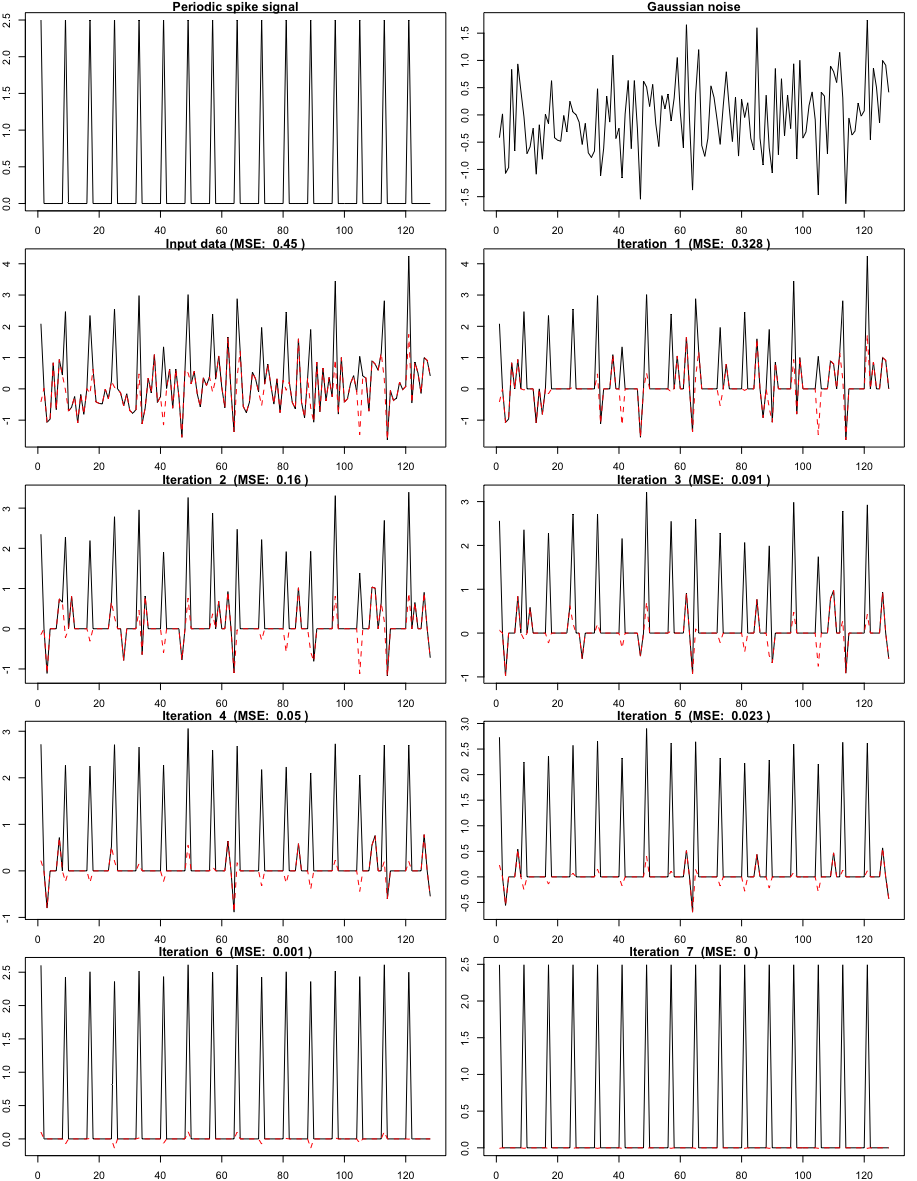}
\end{center}
\caption{Output of the IterativeFT method on $\mathbf{x}$ vector simulated according to the model detailed in Section \ref{sec:vector_denoising}. The top panels show the periodic spike signal and Gaussian noise. The remaining panels show the input data and output from the $h()$ function after each iteration with the error relative to the spike signal captured as a dashed red line and quantified as mean squared error (MSE).}
\label{fig:example}
\end{figure}

\noindent Figure \ref{fig:example} shows an example of $\mathbf{x}$ generated according to this simulation model with $\alpha_{min}=\alpha_{max}=2.5, b=16, \lambda=8$ and $\sigma^2=0.5$. For this specific example, the method converges in seven iterations and perfectly recovers the periodic spike signal. 

\begin{figure}[h]
\begin{center}
\includegraphics[width=0.65\textwidth]{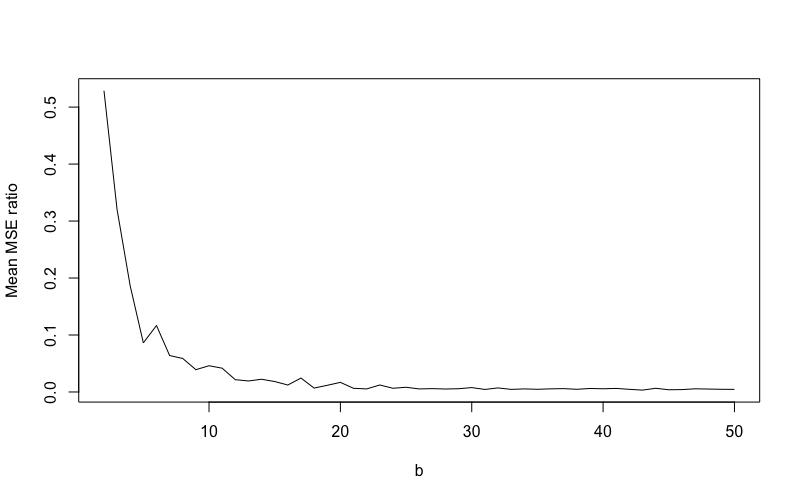}
\end{center}
\caption{Average MSE ratio relative to the number of cycles, b, captured in the input $\mathbf{x}$ vector. The MSE ratio is computed as $MSE_c/MSE_1$ where $MSE_c$ represents the MSE after convergence and $MSE_1$ represents the MSE after the first execution of $h()$ on the input $\mathbf{x}$. }
\label{fig:mse_vs_numcycles}
\end{figure}

To more broadly charaterize signal recovery for this simulation design, multiple $\mathbf{x}$ vectors were generated for different values of $\alpha_{min}, \alpha_{max}, b, \lambda$, and $\sigma^2$. Figure \ref{fig:mse_vs_numcycles} shows the relationship between the MSE ratio achieved on convergence averaged across 50 simulated $\mathbf{x}$ vectors and the number of cycles, $b$, captured in $\mathbf{x}$. The MSE ratio is specifically computed as $MSE_c/MSE_1$ where $MSE_c$ is the mean squared error (MSE) between the output of the method after convergence and the spike signal $\mathbf{s}$ and $MSE_1$ is the MSE for the output from the first execution of $h()$. For this simulation design, the average MSE ratio is very close to 0 for $b \geq 20$, which reflects near perfect recovery of the input periodic spike signal.

\begin{figure}[h]
\begin{center}
\includegraphics[width=0.65\textwidth]{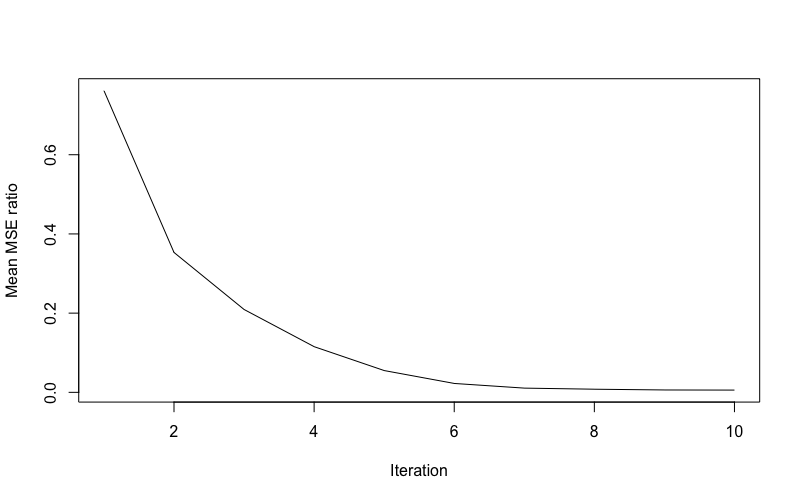}
\end{center}
\caption{Average MSE ratio after each iteration of the algorithm.}
\label{fig:mse_vs_iteration}
\end{figure}

\begin{figure}[h]
\begin{center}
\includegraphics[width=0.75\textwidth]{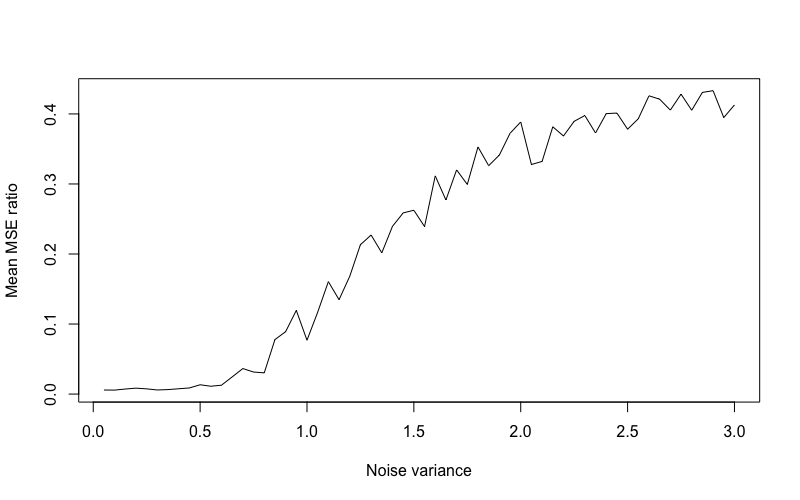}
\end{center}
\caption{Average mean squared error (MSE) ratio relatieve to the variance of the Gaussian noise, $\sigma^2$, added to the periodic spike signal.}
\label{fig:mse_vs_noisevar}
\end{figure}

Figure \ref{fig:mse_vs_iteration} captures the association between the average MSE ratio and the number of iterations completed by the algorithm ($b$ was fixed at 20 for this simulation). These results demonstrate that signal recovery consistently improves on each iteration of the algorithm with the lowest MSE achieved upon convergence. Figure \ref{fig:mse_vs_noisevar} captures the association between the average MSE ratio achieved on convergence and Gaussian noise variance ($b$ was fixed at 20 for this simulation). These results demonstrate the expected increase in signal recovery error with increase noise variance and, importantly, show that the method still achieves improved noise recovery relative to just a single execution of $h()$ at high levels of noise.

\clearpage

\subsection{Detection of spike signal in matrix-valued input}\label{sec:matrix_denoising}

\begin{figure}[h]
\begin{center}
\includegraphics[width=0.5\textwidth]{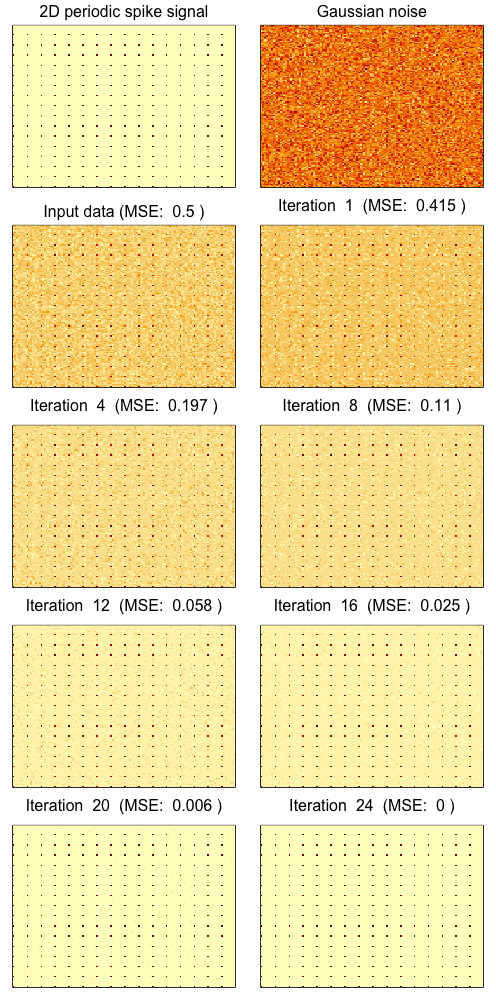}
\end{center}
\caption{Output of the IterativeFT method on a matrix simulated according to the noisy grid design in Section \ref{sec:matrix_denoising}. The top panels show the $\mathbf{S}$ and $\mathcal{E}$ matrices. The remaining panels show the input and output of $h()$ after each iteration with the error relative to $\mathbf{S}$ quantified as MSE.}
\label{fig:matrix_example}
\end{figure}

To explore the generalization where the input is a matrix rather than a vector, we also tested several 2D simulation models. The first model is a noisy $n \times n$ grid $\mathbf{X}$ generated as follows:

\begin{itemize}[labelindent=5pt, topsep=4pt]\setlength{\itemsep}{2pt}
\item Generate a length $n$ vector $\mathbf{s}$ that contains a periodic spike signal using the logic from Section \ref{sec:vector_denoising} (see Figure \ref{fig:example} for visualization of a specific example generated according to this model).
\item Create the signal matrix $\mathbf{S}$ as $\mathbf{S} = \mathbf{s} \mathbf{s}^T$.
\item Generate an $n \times n$ noise matrix $\mathcal{E}$ whose elements are independent $\mathcal{N}(0, \sigma^2)$ random variables.
\item Create the input $n \times n$ matrix $\mathbf{X}$ as $\mathbf{X} = \mathbf{S} + \mathcal{E}$.
\end{itemize}

\noindent Figure \ref{fig:matrix_example} shows an example of $\mathbf{X}$ generated according to this simulation model with $\alpha_{min}=\alpha_{max}=2.5, b=16, \lambda=8$ and $\sigma^2=0.5$. For this specific example, the IterativeFT method converges in 24 iterations and perfectly recovers the periodic spike signal matrix. The convergence properties for this type of matrix input when evaluated across multiple simulations mirror those for the vector input as shown in Figures \ref{fig:mse_vs_numcycles}, \ref{fig:mse_vs_iteration}, and \ref{fig:mse_vs_noisevar}.

For the second example, we simulated a rank 1 matrix $\mathbf{X}$ with a sinusoidal data pattern as follows:

\begin{itemize}[labelindent=5pt, topsep=4pt]\setlength{\itemsep}{2pt}
\item Create a length $100$ vector $t$ as 100 uniformly spaced points between 0 and 1.
\item Generate a length $100$ vector $\mathbf{x}$ that contains the sinusoidal signal $sin(3 \pi t^2)^2$.
\item Generate a length $100$ vector $\mathbf{y}$ that contains the sinusoidal signal $sin(\pi t)^2$.
\item Create the signal matrix $\mathbf{S}$ as $\mathbf{S} = \mathbf{x} \mathbf{y}^T$.
\item Generate an $100 \times 100$ noise matrix $\mathcal{E}$ whose elements are independent $\mathcal{N}(0, 0.1)$ random variables.
\item Create the input $100 \times 100$ matrix $\mathbf{X}$ as $\mathbf{X} = \mathbf{S} + \mathcal{E}$.
\end{itemize}

\begin{figure}[h]
\begin{center}
\includegraphics[width=1\textwidth]{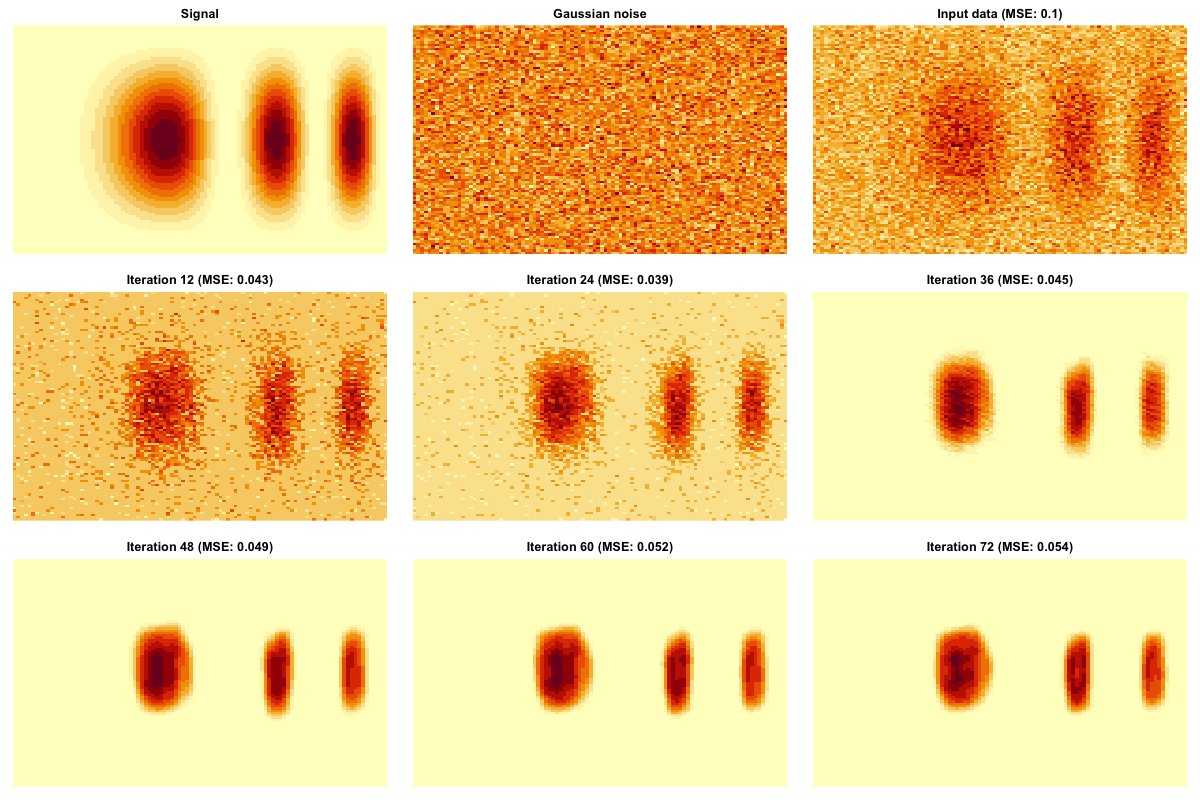}
\end{center}
\caption{Output of the IterativeFT method on a matrix simulated according to rank 1 sinusoidal model in Section \ref{sec:matrix_denoising}. The top panels show the $\mathbf{S}$, $\mathcal{E}$, and $\mathbf{X}$ matrices. The remaining panels show the input and output of $h()$ after different iterations with the error relative to $\mathbf{S}$ quantified as MSE.}
\label{fig:sin_matrix_example}
\end{figure}
 
\noindent Figure \ref{fig:sin_matrix_example} shows an example of $\mathbf{X}$ generated according to this sinusoidal simulation model. For this specific example, the IterativeFT method converges in 72 iterations and recovers a sparse version of the sinusoidal input pattern.

\clearpage

\subsection{Analysis of spike signal detection}\label{sec:analysis}

The simulation-based results displayed in Sections \ref{sec:vector_denoising} and \ref{sec:matrix_denoising} can be understood as follows:

\begin{itemize}[labelindent=5pt, topsep=4pt]\setlength{\itemsep}{2pt}
\item The elements of the generated input vector $\mathbf{x}$ will be stochastically larger for elements that are non-zero in the periodic signal vector $\mathbf{s}$ given that the expected value of $\mathbf{x}$ is based solely on $\mathbf{s}$:
\begin{align*}
E[\mathbf{x}[j]] &= E[\mathbf{s}[j] + \varepsilon[j]] \\
E[\mathbf{x}[j]] &= E[\mathbf{s}[j]] + E[\varepsilon[j]] & \text{E of sum is sum of E} \\
E[\mathbf{x}[j]] &= E[\mathbf{s}[j]] & \text{$\varepsilon[j] \sim \mathcal{N}(0,1)$ so $E[\varepsilon[j]] = 0$}
\end{align*}
\item These non-zero signal vector elements will therefore be less likely to be set to 0 by the sparsification version of $h()$, 
which results in the frequency spectra of $h(\mathbf{x})$ being more strongly based on the frequency spectra of $\mathbf{s}$ than the spectra of $\boldsymbol{\varepsilon}$.
\item These characteristics of the frequency spectra of the output from $h(\mathbf{x})$ mean that the output of $dft(h(\mathbf{x}))$ will be stochastically larger for elements that correspond to the spectra of $\mathbf{s}$.
\item The sparsification version of $g()$ will therefore be more likely to retain frequency components associated with $\mathbf{s}$ and set to 0 frequency components corresponding to $\boldsymbol{\varepsilon}$.
\item Repeated iterations will reinforce these patterns until a stable sparsity structure is obtained in $\mathbf{x}$, which will tend to correspond to the very sparse frequency spectra of $\mathbf{s}$.
\end{itemize}

\noindent To understand why the iterative application of both $h()$ and $g()$ on real and frequency representations of the data is needed one can consider scenarios where only one of $h()$ or $g()$ is performed (as outlined in Section \ref{sec:trivial}, such scenarios are inherently non-iterative):
\begin{itemize}[labelindent=5pt, topsep=4pt]\setlength{\itemsep}{2pt}
\item Only $h()$ is executed: Although a single execution of the sparsification version of $h()$ will tend to reduce the error between $\mathbf{s}$ and $\mathbf{x}$, noise of a sufficient amplitude will result in the incorrect removal of some parts of $\mathbf{s}$ (i.e., negative noise may lower the value of a spike below the threshold) and incorrect retention of some parts of $\boldsymbol{\varepsilon}$ (i.e., the amplitude of the random noise may be larger than the threshold), e.g.  Figure \ref{fig:example}. The performance of a single $h()$ execution is assessed more comprehensively in \ref{sec:comp_eval}.
\item Only $g()$ is executed: In this scenario, only a single execution of $dft()$ is performed on $\mathbf{x}$, followed by execution of $g()$ before finally transforming back via $dft^{-1}()$, i.e., the output is given by $dft^{-1}(g(dft(\mathbf{x})))$. If one had prior knowledge of the spectra of $\mathbf{s}$, then it would be possible specifically target the desired frequencies via $g()$, e.g., a bandpass filter (though as shown in Section \ref{sec:comp_eval} below, even prior knowledge of an appropriate frequency band may not yield good denoising performance). Lacking this prior knowledge, however, sparsification via $g()$ can only be based the magnitude of frequency domain coefficients and, while it will tend to preserve the portions of the spectra of $\mathbf{x}$ associated with $\mathbf{s}$, random noise will cause the incorrect removal of spectral components that are due to $\mathbf{s}$ and the incorrect retention of spectral components due to $\boldsymbol{\varepsilon}$. The performance of a single $g()$ execution is assessed more comprehensively in \ref{sec:comp_eval}.
\end{itemize}

To understand the simulation-based results displayed in Figures \ref{fig:convergence_vs_n}, \ref{fig:convergence_vs_p}, \ref{fig:example}, \ref{fig:mse_vs_numcycles}, \ref{fig:mse_vs_iteration}, and \ref{fig:mse_vs_noisevar}, we can evaluate the mathematical representation of one iteration of the algorithm given by \eqref{eqn:iteration_rep} for $k \geq 1$ when $h()$ and $g()$ induce sparsity. To restate \eqref{eqn:iteration_rep}:

%
%
%
%
%
%
%
%


\section{Comparative evaluation of denoising performance}\label{sec:comp_eval}

To evaluate the performance of the IterativeFT method relative to existing signal denoising techniques, a simulation study was undertaken on data generated according to several spike signal patterns with varying signal-to-noise ratios and spike frequencies.

\subsection{Simulation design}\label{sec:sim_design}

The input data was generated as a length $n$ vector $\mathbf{x} = \mathbf{s} + \boldsymbol{\varepsilon}$ where the elements of $\boldsymbol{\varepsilon}$ were generated as independent $\mathcal{N}(0,\sigma^2)$ random variables and the signal $\mathbf{s}$ followed one of three different periodic spike signal patterns:

\begin{enumerate}
\item \textbf{Uniform spike}: Elements $s_i, i \in (1,...,n)$ of $\mathbf{s}$ are set to 0 for all $i \neq a\lambda$ and set to the constant $\delta$ for $i = a\lambda, a \in (1,...,b)$ where $b=floor(n/\lambda)$ is the total number of cycles and $\lambda \geq 2$ is the period. The first panel in Figure \ref{fig:uniform_example} illustrates this signal pattern, which was also used for the example in Section \ref{sec:vector_denoising} above.
\item \textbf{Alternating sign spike}: The sparsity pattern is similar to the uniform spike model above but the non-zero values alternate between $\delta$ and $-\delta$. The first panel in Figure \ref{fig:alt_sign_example} below illustrates this signal pattern.
\item \textbf{Alternating size spike}: The sparsity pattern is similar to the uniform spike model above but the non-zero values alternate between $\delta/2$ and $\sqrt{7/4}\delta$ (these values yield a signal power equivalent to that for a signal with $\delta$ sized spikes).
The first panel in Figure \ref{fig:alt_size_example} below illustrates this signal pattern.
\end{enumerate}

\noindent For the simulation results presented in Sections \ref{sec:simple_results}-\ref{sec:alt_size_results} below, $n=1024$ and $\delta=2$. The noise variance, $\sigma^2$, was set to yield a signal-to-noise ratio ranging between 0.05 and 2. The $n$ value of 1024 corresponds to a Nyquist frequency of 512 Hz and the spike period $\lambda$ was set to yield a ratio of spike frequency to Nyquist frequency ranging between 0.0625 (for $\lambda=32$) and 1 (for $\lambda=2$). For the single examples in Figures \ref{fig:uniform_example}, \ref{fig:alt_sign_example} and \ref{fig:alt_size_example}, $n=128$, $\lambda=8$ (i.e., a ratio of spike frequency to Nyquist frequency of 0.25) and $\sigma^2$ was set to yield a signal-to-noise ratio of 1.

\subsection{Comparison methods}\label{sec:comp_methods}

We compared the denoising performance of the IterativeFT method (Algorithm 1 using the $h()$, $g()$ and $c()$ functions defined in Section \ref{sec:denoising}) against four other techniques:

\begin{itemize}
\item \textbf{Real domain thresholding}: This method performs a single thresholding operation on $\mathbf{x}$ using $h()$.
\item \textbf{Frequency domain thresholding}: This method performs a single frequency domain thresholding operation using $g()$, i.e., the denoised output is generated as $dft^{-1}(g(dft(x)))$.
\item \textbf{Butterworth bandpass filtering}: This method is realized by a Butterworth \cite{Butterworth1930} 3-order bandpass filter (as implemented by the \textit{butter()} function in v0.3-5 of the \textit{gsignal} R package \cite{Van-Boxtel-G.J.M.-et-al.:2021aa}) where the band is set to the spike signal frequency (as a fraction of the Nyquist frequency) $\pm 0.05$, e.g, if the spike signal is 0.25 of the Nyquist frequency, the band is set to 0.2 to 0.3.
\item \textbf{Wavelet filtering}: This method is based on the usage example for the \textit{threshold.wd()} function in v4.7.4 of the \textit{wavethresh} \cite{wavethresh} R package. Specifically, a discrete wavelet transform is first applied to $\mathbf{x}$ using the Daubechies least-asymmetric orthonormal compactly supported wavelet with 10 vanishing moments and interval boundary conditions \cite{COHEN199354}. This wavelet transform is realized using the \textit{wd()} function in the \textit{wavethresh} R package with \textit{bc="interval"} and the default wavelet family and smoothness settings. The wavelet coefficients at all levels are then thresholded using a threshold value computed according to a universal policy and madmad deviance estimate on the finest coefficients followed by application of an inverse wavelet transformation.
\end{itemize}

\subsection{Results for uniform spike model}\label{sec:simple_results}

\begin{figure}[h]
\begin{center}
\includegraphics[width=0.8\textwidth]{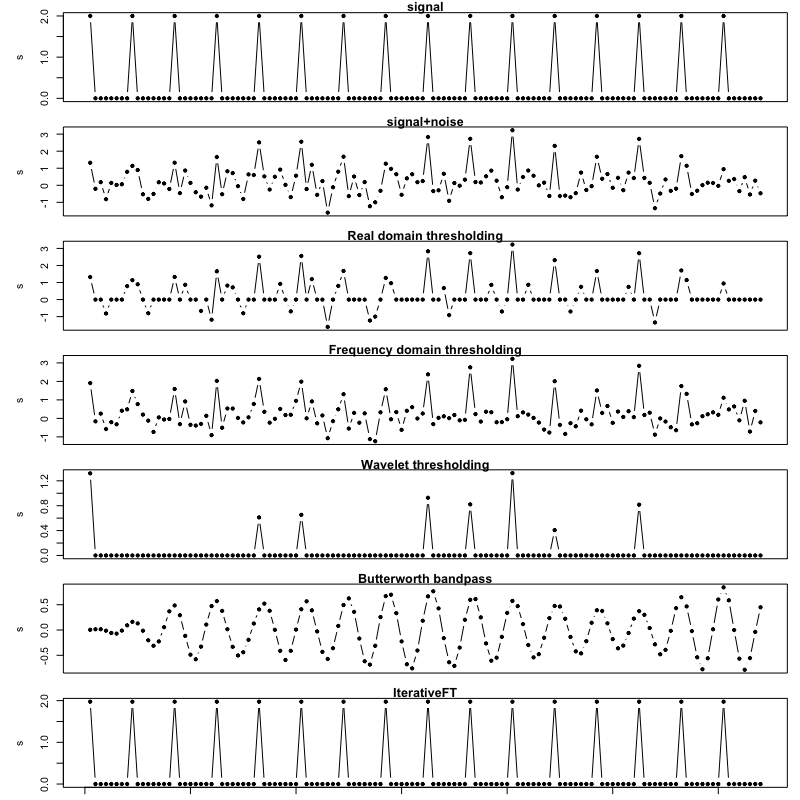}
\end{center}
\caption{Output of the IterativeFT method and comparison denoising methods on a single vector generated according to the uniform spike model.}
\label{fig:uniform_example}
\end{figure}

\begin{figure}[h]
\begin{center}
\includegraphics[width=0.8\textwidth]{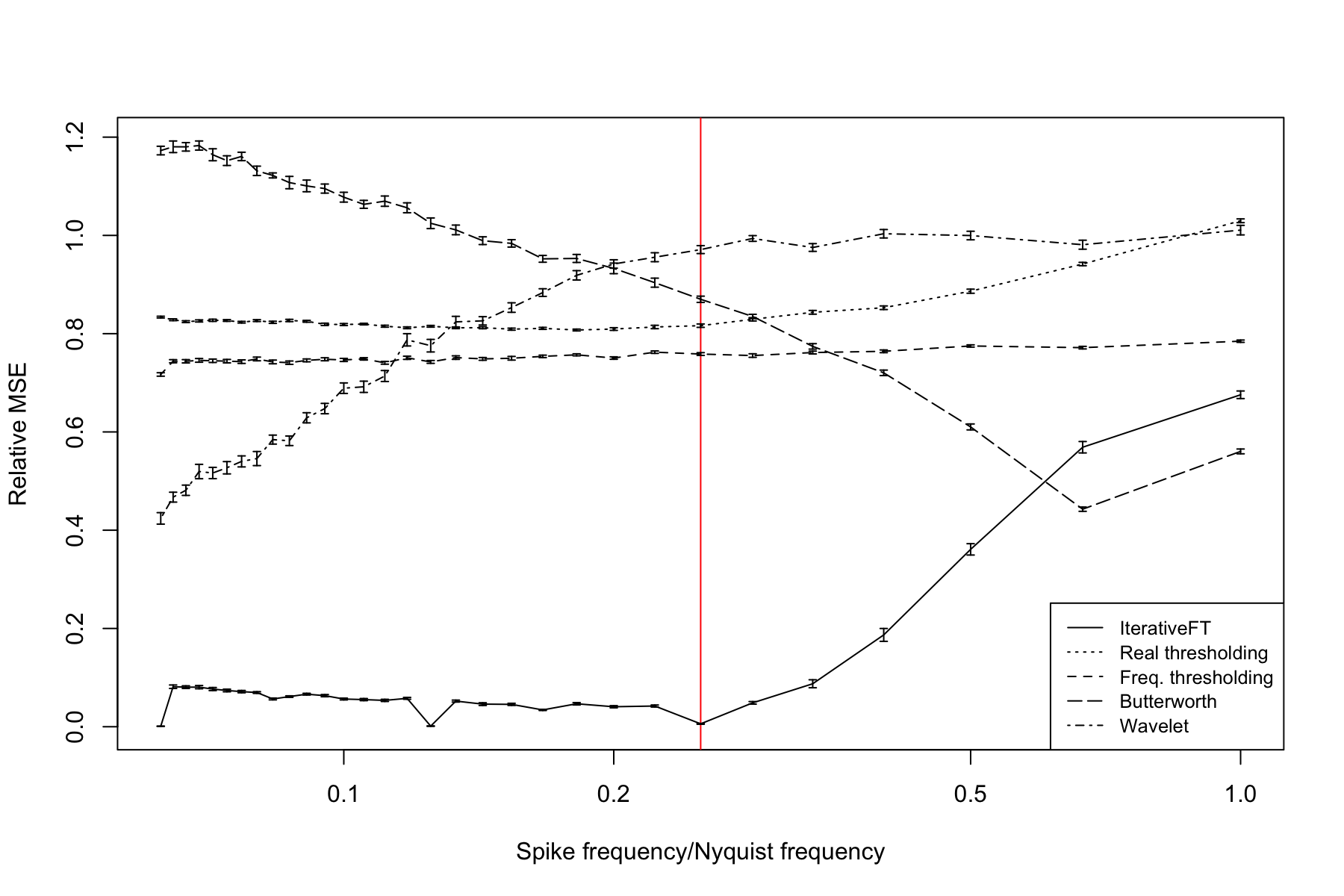}
\end{center}
\caption{Signal denoising performance for the uniform spike model relative to spike signal frequency. For each frequency value (shown as a fraction of the Nyquist frequency), 25 vectors were simulated according to the uniform spike model and signal denoising was performed using the five evaluated techniques. Performance was assessed as the mean MSE between the denoised data and the underlying signal, which is plotted relative to that MSE measured on the undenoised data. Error bars are $\pm$ 1 SE.
The vertical red line represents the relative frequency value used in the simulations shown in Figure \ref{fig:uniform_snr}.}
\label{fig:uniform_freq}
\end{figure}

\begin{figure}[h]
\begin{center}
\includegraphics[width=0.8\textwidth]{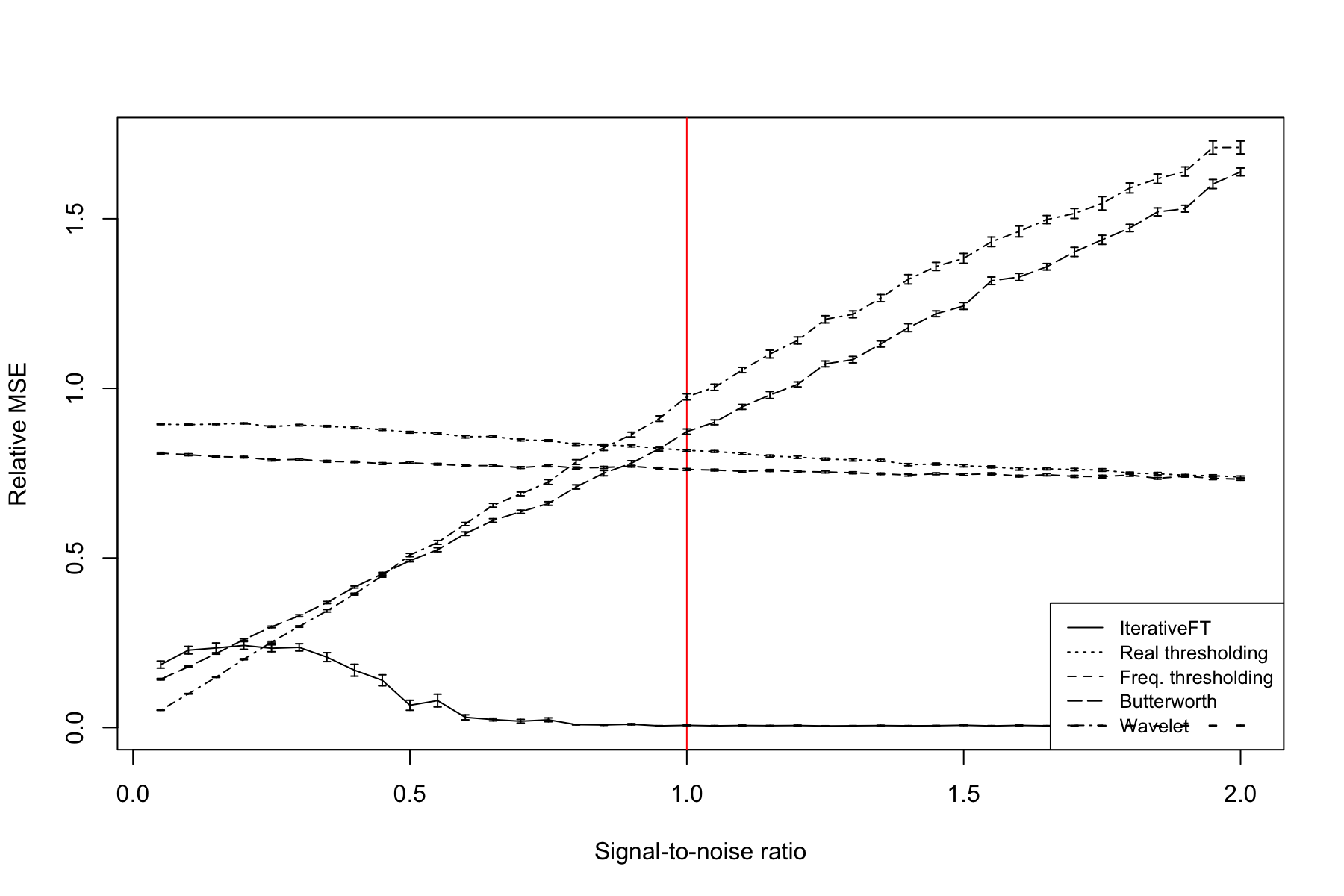}
\end{center}
\caption{Signal denoising performance for the uniform spike model relative to signal-to-noise ratio. For each signal-to-noise ratio value, 25 vectors were simulated according to the uniform spike model and signal denoising was performed using the five evaluated techniques. Performance was assessed as the mean MSE between the denoised data and the underlying signal, which is plotted relative to that MSE measured on the undenoised data. Error bars are $\pm$ 1 SE.
The vertical red line represents the signal-to-noise value used in the simulations shown in Figure \ref{fig:uniform_freq}.}
\label{fig:uniform_snr}
\end{figure}

Figures \ref{fig:uniform_example}-\ref{fig:uniform_snr} illustrate the comparative denoising performance for the evaluated techniques on data simulated according to the uniform spike model. Figure \ref{fig:uniform_example} shows the outputs for a single example vector with a signal-to-noise (SNR) ratio of 1 and ratio of spike signal frequency to Nyquist frequency of 0.25. For this example, the IterativeFT method is able to perfectly recover the spike signal. In contrast, the other evaluated techniques have only marginal signal recovery performance.

Figure \ref{fig:uniform_freq} shows the relationship between denoising performance and spike signal frequency at a signal-to-noise (SNR) value of 1. Denoising performance is plotted on the y axis as the mean MSE across 25 simulated inputs between the denoised data and the underlying signal and is plotted relative to the MSE measured on the undenoised data. A relative MSE value equal to 1 indicates null performance, i.e., the denoising method is equivalent to leaving the data unchanged; values above 1 indicate that the method further corrupts the signal. For this simulation model, the IterativeFT technique offers dramatically better denoising performance than the other four techniques with the exception of the high frequency domain where the Butterworth bandpass filter has slightly better performance. The simple real and frequency domain thresholding methods, i.e., a single application of either the $h()$ or $g()$ sparsification functions, are slightly better than null and relatively insensitive to signal frequency. The wavelet filter offers the good performance at low frequency values with performance decreasing towards the null level as the signal frequency increases. By contrast, the Butterworth filter is worse than null at low frequency values with performance steadily improving as frequency increases. The three notable dips in relative MSE for the IterativeFT method correspond to spike periods of 32, 16 and 8, which are all factors of 1024, the length of the input vector.

Figure \ref{fig:uniform_snr} illustrates the relationship between denoising performance and the SNR value at a spike frequency that is 0.25 of the Nyquist frequency.
Similar to Figure \ref{fig:uniform_freq}, both of the simple thresholding methods have performance that is only marginally better than null but, as expected, does improve slightly increasing SNR. The IterativeFT method is substantially better than all comparison techniques at SNR values above 0.5. At low SNR values, performance of the IterativeFT method decreases and falls below both the wavelet and Butterworth filtering techniques at SNR values below $\sim$0.25. The relative performance of the wavelet and Butterworth methods shows a steady decline with increasing SNR, which is surprising the error should decrease as SNR increases. This unexpected trend is due to the fact that the relative rather than absolute MSE is shown. In particular, the absolute MSE for these methods does decrease as the SNR increases, however, the MSE relative to no modifications increases with both filtering methods giving worse than null performance at SNR values greater than $\sim$ 1.1.

\clearpage

\subsection{Results for alternating sign spike model}\label{sec:alt_sign_results}

\begin{figure}[h]
\begin{center}
\includegraphics[width=0.8\textwidth]{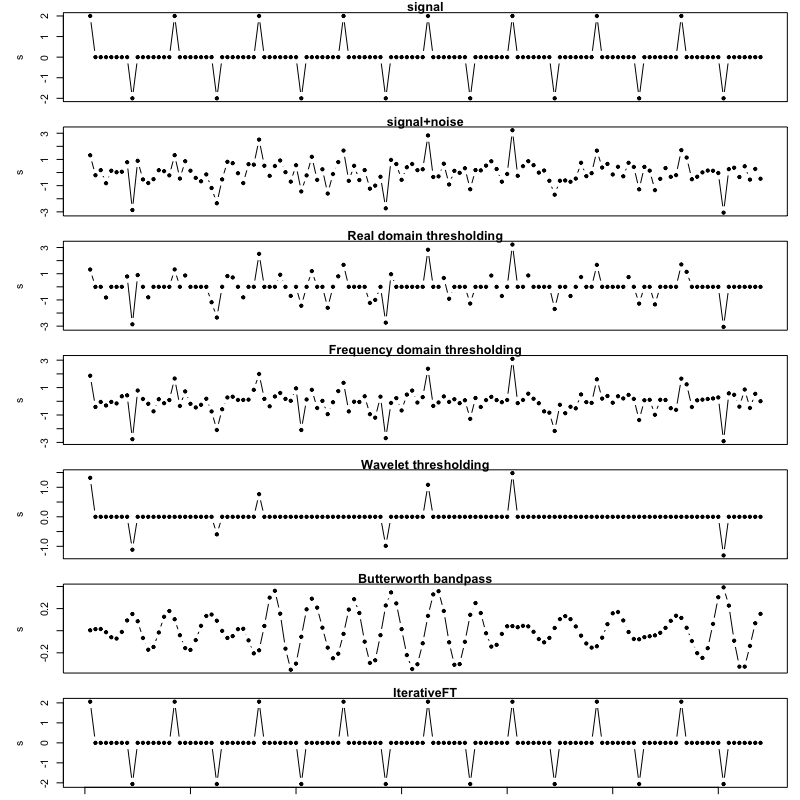}
\end{center}
\caption{Output of the IterativeFT method and comparison denoising methods on a single vector generated according to the alternating sign spike model.}
\label{fig:alt_sign_example}
\end{figure}

\begin{figure}[h]
\begin{center}
\includegraphics[width=0.8\textwidth]{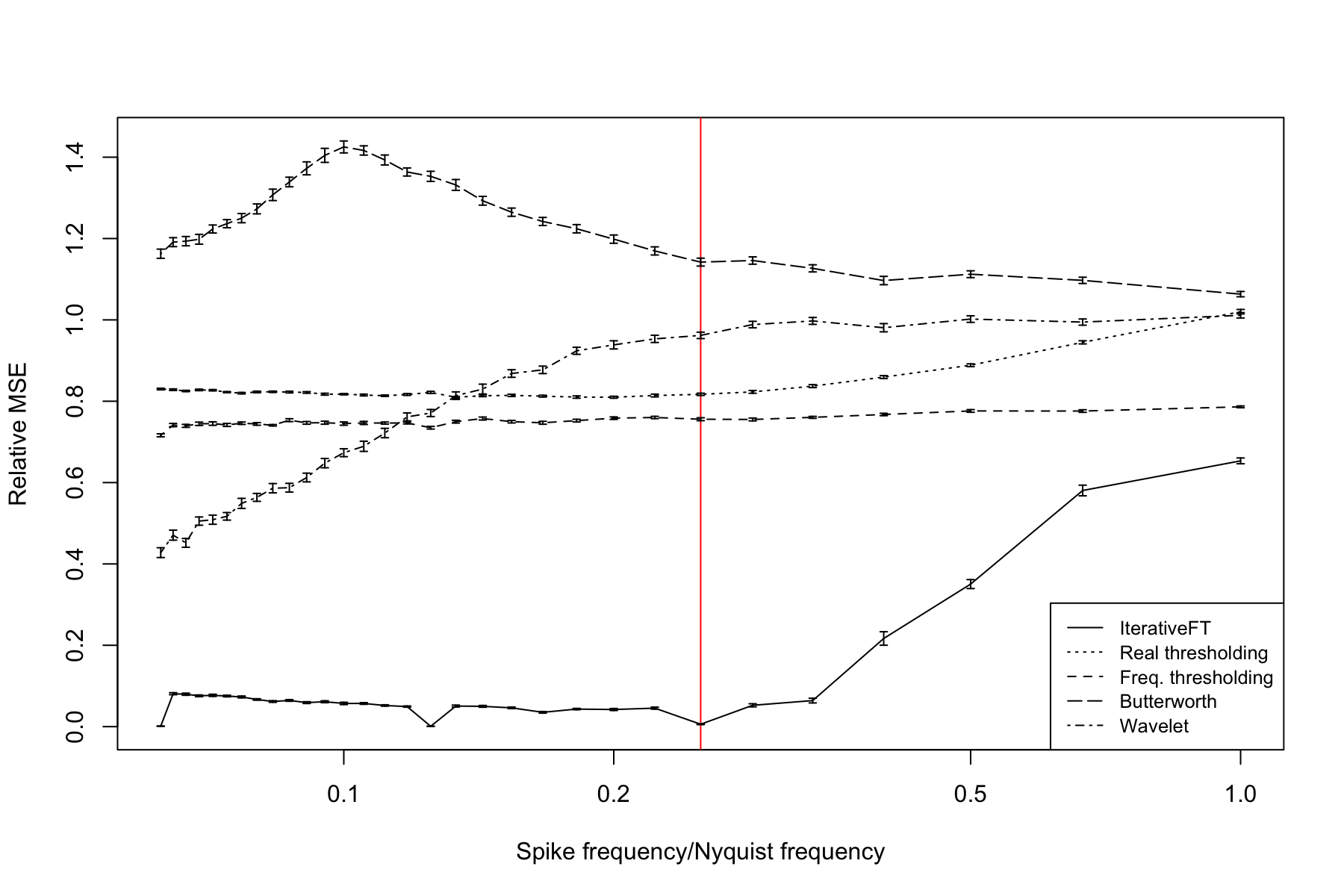}
\end{center}
\caption{Signal denoising performance for the alternating sign spike model relative to spike signal frequency. Plot interpretation follows that for Figure \ref{fig:uniform_freq}.}
\label{fig:alt_sign_freq}
\end{figure}

\begin{figure}[h]
\begin{center}
\includegraphics[width=0.8\textwidth]{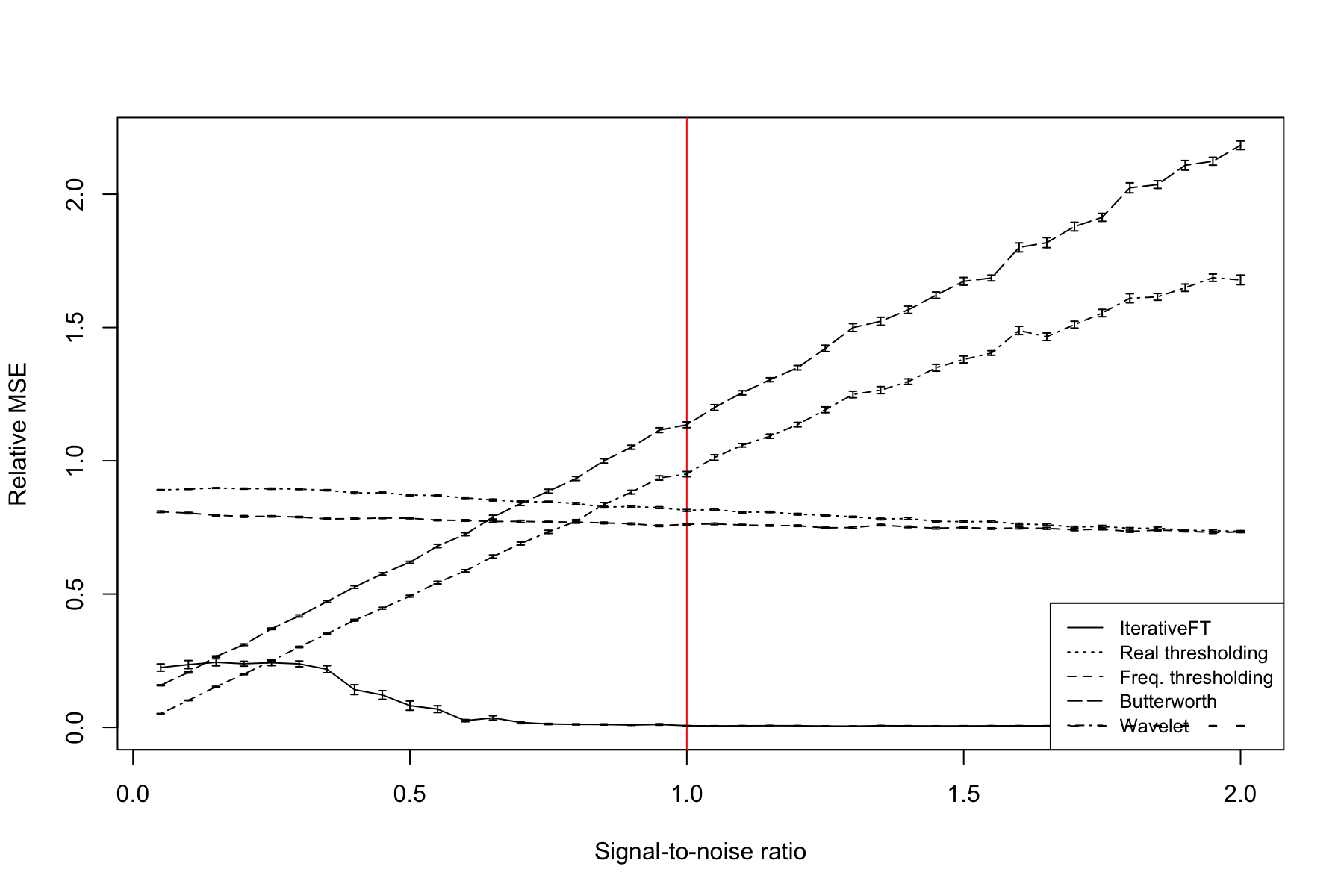}
\end{center}
\caption{Signal denoising performance for the alternating sign spike model relative to signal-to-noise ratio.
Plot interpretation follows that for Figure \ref{fig:uniform_snr}.}
\label{fig:alt_sign_snr}
\end{figure}

Figures \ref{fig:alt_sign_example}-\ref{fig:alt_sign_snr} illustrate comparative denoising performance on data simulated according to the alternating sign spike model. Figure \ref{fig:alt_sign_example} shows the outputs for a single example vector with a signal-to-noise (SNR) ratio of 1 and ratio of spike signal frequency to Nyquist frequency of 0.25. Similar to the uniform spike example in Figure \ref{fig:uniform_example}, the IterativeFT method perfectly recovers the spike signal with poor performance by the comparative techniques.

Figures \ref{fig:alt_sign_freq} and \ref{fig:alt_sign_snr} visualize the performance of the evaluated methods on the alternating sign model relative to signal frequency and SNR. The results are generally similar to those for the uniform spike model with the exception that the Butterworth bandpass filter has worse than null performance across all tested signal frequencies.

\clearpage

\subsection{Results for alternating size spike model}\label{sec:alt_size_results}

\begin{figure}[h]
\begin{center}
\includegraphics[width=0.8\textwidth]{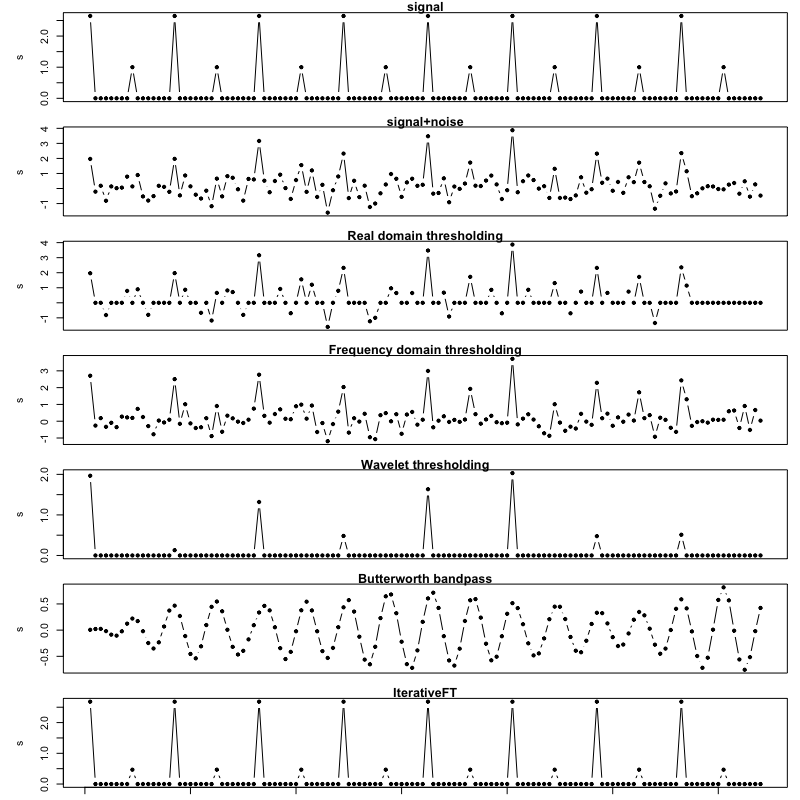}
\end{center}
\caption{Output of the IterativeFT method and comparison denoising methods on a single vector generated according to the alternating size spike model.}
\label{fig:alt_size_example}
\end{figure}

\begin{figure}[h]
\begin{center}
\includegraphics[width=0.8\textwidth]{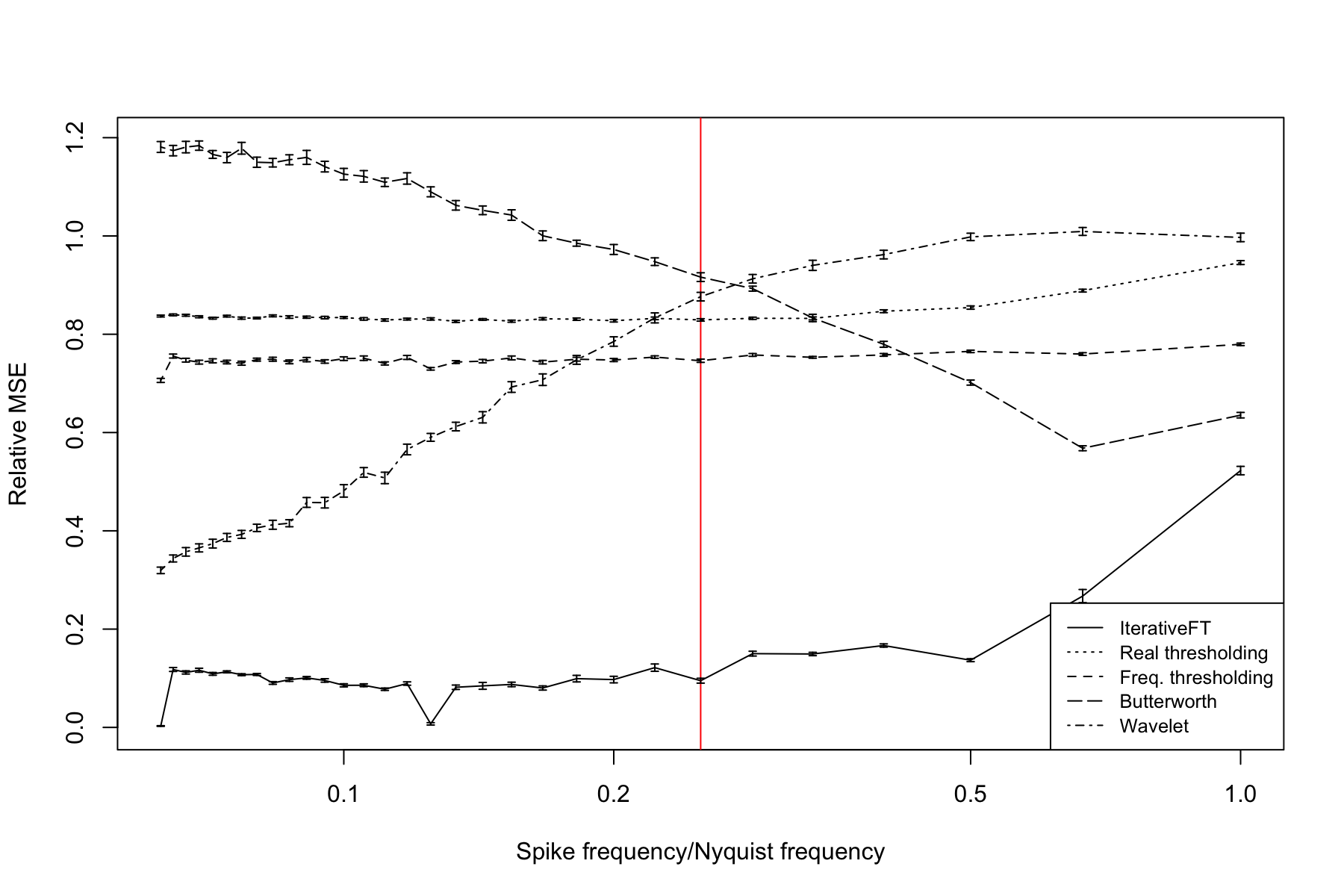}
\end{center}
\caption{Signal denoising performance for the alternating size spike model relative to spike signal frequency. 
Plot interpretation follows that for Figure \ref{fig:uniform_freq}.}
\label{fig:alt_size_freq}
\end{figure}

\begin{figure}[h]
\begin{center}
\includegraphics[width=0.8\textwidth]{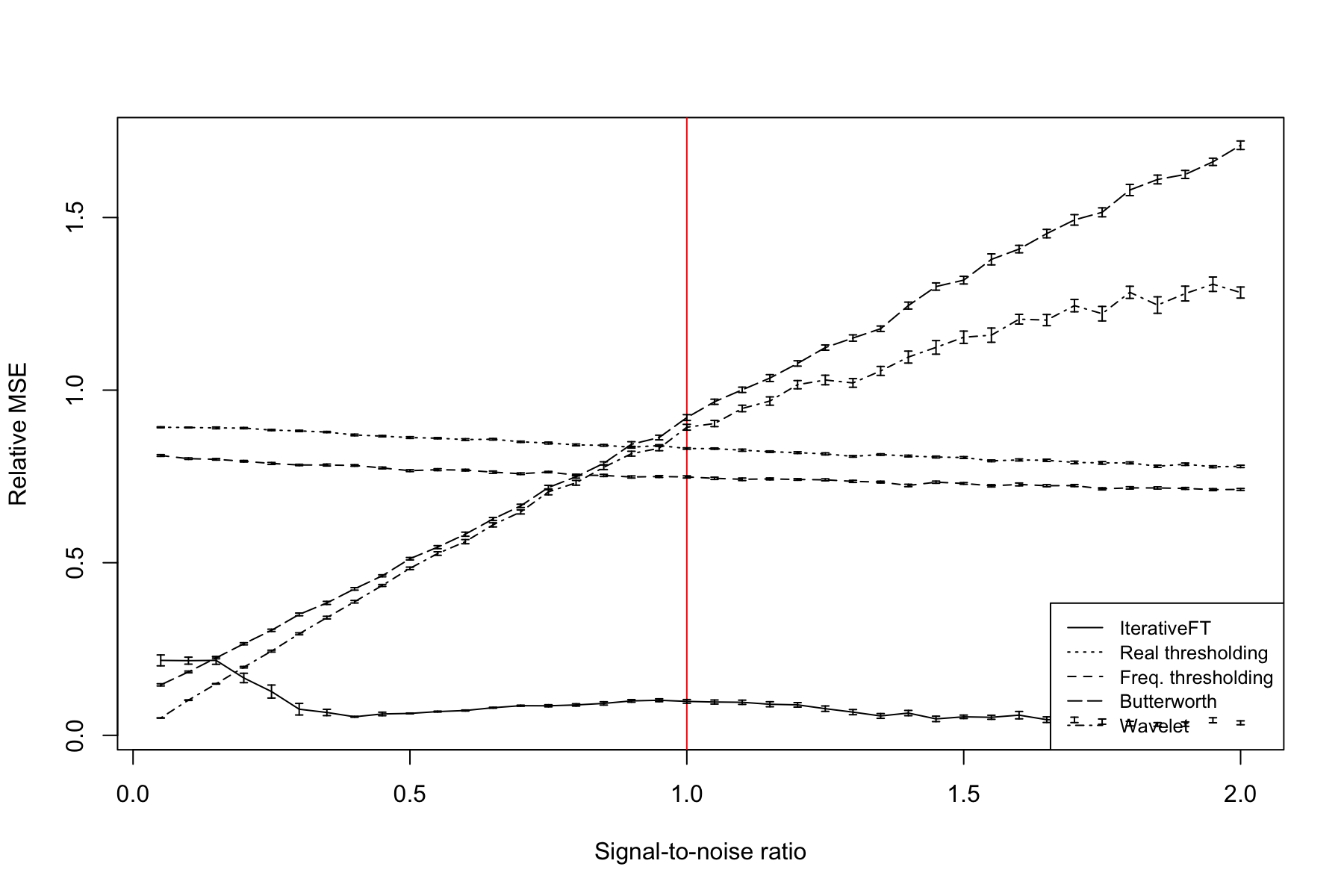}
\end{center}
\caption{Signal denoising performance for the alternating size spike model relative to signal-to-noise ratio. 
Plot interpretation follows that for Figure \ref{fig:uniform_snr}.}
\label{fig:alt_size_snr}
\end{figure}

Figures \ref{fig:alt_size_example}-\ref{fig:alt_size_snr} illustrate comparative denoising performance on data simulated according to the alternating size spike model. Figure \ref{fig:alt_size_example} shows the outputs for a single example vector with a signal-to-noise (SNR) ratio of 1 and ratio of spike signal frequency to Nyquist frequency of 0.25. Similar to the uniform spike and alternating sign examples, the IterativeFT method provides excellent signal recovery though it does underestimate the magnitude of the smaller spike; the other evaluated methods are worse by comparison.

Figures \ref{fig:alt_size_freq} and \ref{fig:alt_size_snr} visualize the performance of the evaluated methods on the alternating size model relative to signal frequency and SNR. The results are generally similar to those for the uniform spike model with the exception that performance for the Butterworth bandpass filter is never better than the IterativeFT method.

\clearpage

\section{Conclusion}

In this paper we explored a family of iterative algorithms based on the repeated execution of discrete and inverse discrete Fourier transforms on real valued vector or matrix data. One interesting member of this family, which we refer to as the IterativeFT method, is motivated by the discrete Fourier transform uncertainty principle and involves the application of a thresholding operation to both the real domain and frequency domain data with convergence obtained when real domain sparsity hits a stable pattern. As we demonstrated through simulation studies, the IterativeFT method can effectively recover periodic spike signals across a wide range of spike signal frequencies and signal-to-noise ratios. Importantly, the performance of the IterativeFT method is significantly better than standard non-iterative denoising techniques including real and frequency domain thresholding, wavelet filtering and Butterworth bandpass filtering. An R package implementing this technique and related resources can be found at \href{https://hrfrost.host.dartmouth.edu/IterativeFT}{https://hrfrost.host.dartmouth.edu/IterativeFT}. Areas for future work include on this method include:
\begin{itemize}[labelindent=5pt, topsep=4pt]\setlength{\itemsep}{2pt}
\item Expanding the comparative evaluation to include other denoising techniques, e.g., basis pursuit \cite{471413}, and a broader range of periodic signal and noise models, e.g., composition of spike signals at multiple frequencies/amplitudes, mixtures of harmonic and spike signals, etc.
\item Exploring other classes of $h()$, $g()$ and $c()$ functions and associated analysis problems, e.g., soft thresholding.
\item Exploring the generalization of the IterativeFT method to complex or hypercomplex-valued inputs.
\item Exploring the generalization of the IterativeFT method to other invertable discrete transforms, e.g., discrete wavelet transform \cite{10.5555/1525499}.
\end{itemize}

\section*{Acknowledgments}

This work was funded by National Institutes of Health grants R35GM146586 and R21CA253408. We would like to thank Anne Gelb and Aditya Viswanathan for the helpful discussion.

\bibliographystyle{unsrt}

\end{document}